\documentclass[aps,prl,
reprint,superscriptaddress,nofootinbib,floatfix]{revtex4-1}

\usepackage{lmodern}
\usepackage{amsmath,amssymb,amsfonts,mathtools,bm}
\usepackage{graphicx}
\usepackage[colorlinks=true,linkcolor=blue,citecolor=blue,urlcolor=blue,plainpages=false,pdfpagelabels]{hyperref}

\newtheorem{theorem}{Theorem}

\DeclareMathOperator{\Tr}{Tr}

\newcommand{\ii}{\mathrm{i}}
\newcommand{\dd}{\mathrm{d}}
\newcommand{\one}{\mathbb{I}}
\newcommand{\cH}{\mathcal{H}}

\newcommand{\cO}{\mathcal{O}}

\newcommand{\Cred}{\mathcal{C}_{\rm red}}
\newcommand{\mrec}{m_{\rm rec}}
\newcommand{\sqdd}{\Delta_{\rm SQD}}
\newcommand{\barF}{\overline F}
\newcommand{\DeltaPi}{\Delta_\Pi^S}
\newcommand{\eps}{\varepsilon}
\newcommand{\rhozero}{\rho^{0}}
\newcommand{\rhoact}{\rho^{\rm act}}
\graphicspath{{figures/}}

\begin{document}

\title{Hamiltonian Thresholds for Objective Records}

\author{Jie Gu}
\email{jiegu1989@gmail.com}
\affiliation{Chengdu Academy of Education Sciences, Chengdu 610036, China}

\date{\today}

\begin{abstract}
Classical objectivity requires an environment to witness a pointer value, not merely to decohere it.  We derive microscopic Hamiltonian laws for this distinction.  Exact QND product monitoring yields a certified two-edge strong-Darwinism window: sufficient witnesses cross the optimal record-discrimination edge while retaining an unobserved complement that crosses the residual-coherence edge.  Under additional regularity assumptions, the supplement gives converse bounds at the same logarithmic scale.  For controlled-phase monitors, environmental asymmetry gives a single-fragment information bound, a redundancy converse, and, in efficient i.i.d. streams, an achievable block-capacity law for redundant witnesses.  A finite-temperature spin-phase bath shows the physical separation: hot probes can decohere efficiently while losing the local asymmetry needed for objective witnesses.
\end{abstract}

\maketitle

\paragraph{Introduction.---}
Macroscopic reality is normally learned indirectly.  We do not interrogate the
center-of-mass state of a dust grain, a spin, or a measuring apparatus itself;
we intercept photons, phonons, scattered molecules, or other environmental
carriers that have already interacted with it.  The quantum-to-classical problem
therefore has two logically distinct parts.  Decoherence explains why
interference between preferred pointer states becomes locally invisible in the
reduced state of the system \cite{ZurekPhysToday,ZurekRMP,SchlosshauerBook}.
Objectivity requires more: many independently accessible fragments of the
environment must carry the same classical record, so that different observers
can infer the same pointer value without perturbing the system or one another
\cite{OllivierPoulinZurek,OllivierPoulinZurek2005,BlumeKohoutZurek2005,
BlumeKohoutZurek,ZurekNatPhys,RiedelZurek2010}.

This distinction is not merely semantic.  A bath can decohere a system very
efficiently while failing to provide locally readable records.  Conversely, a
fragment may be strongly correlated with the system but still owe part of its
mutual information to quantum correlations that are not objective records.
The usual Darwinistic signature, a plateau of mutual information versus
fragment size, therefore does not by itself certify classical objectivity
\cite{ZwolakZurek2013,Korbicz2021}.  Recent work has sharpened this warning by
separating redundancy from consensus, by quantifying when independently
accessed fragments support observer agreement, by showing that non-averaged
mutual information can be misleading in inhomogeneous environments, by recasting
emergent objectivity as a metrological information-acquisition problem, and by
identifying branching structure as the low-discord state structure compatible
with Darwinistic classicality \cite{ChisholmInnocentiPalma2023,
TouilYanZurek2025,ChisholmInnocentiPalma2024,
KielyChisholmTouilDeffnerLandiCampbell2026,
TouilAnzaDeffnerCrutchfield2024}.  Spectrum-broadcast-structure and strong
quantum Darwinism make the missing requirement explicit: an objective fragment
must contain the full pointer Holevo information and have negligible
pointer-basis discord with the system \cite{KorbiczHorodeckiHorodecki2014,
HorodeckiKorbiczHorodecki2015,LeOlaya}.  This is also the operational content
of local broadcasting: only classical correlations can be redundantly and
locally proliferated \cite{PianiHorodeckiHorodecki2008,
BrandaoPianiHorodecki2015}.

The microscopic question is then sharper than ``does the environment decohere
the system?''  It is: which Hamiltonian resources make a fragment a witness,
and how many such witnesses can coexist?  This question has become especially
timely because recent work has moved beyond state-level diagnostics toward
amplification bounds, consensus measures, metrological witnesses, encoding
transitions, von-Neumann measurement models, Hamiltonian classifications, and
thermalization-based mechanisms for Darwinistic redundancy
\cite{TouilYanGirolamiDeffnerZurek2022,Zwolak2022,TouilYanZurek2025,
KielyChisholmTouilDeffnerLandiCampbell2026,FerteCao2024,DoucetDeffner2024,
AcevedoWehr2024,CaoNussinov2026}.  This Letter answers the
finite-fragment version of that question for repeated controlled monitoring.
We show that strong quantum Darwinism is governed by two competing Hamiltonian
exponents.  An observed fragment is certified as an objective witness when it
crosses a record-discrimination edge, controlled by quantum Chernoff information
\cite{Helstrom,AudenaertChernoff,NussbaumSzkola,ZwolakRiedelZurek2014,
TouilYanGirolamiDeffnerZurek2022,Zwolak2022}, while the unobserved complement
still crosses a residual-decoherence edge that removes pointer coherences from
the joint state of the system and fragment.  Under the regular finite-model
assumptions stated in the supplement, failure to cross the same edges gives a
matching logarithmic converse.  Thus the certified observer window is finite:
fragments that are too small are unreadable, whereas fragments that are too
large leave insufficient environment to classicalize their correlation with the
system.

We further identify the resource behind this window for controlled-phase
monitors.  Decoherence can arise from phase dispersion, but readable local
records require environmental asymmetry relative to the coupling generator,
closely related to coherence-asymmetry resources \cite{BaumgratzCoherence,
MarvianSpekkens}.  The resulting asymmetry bound is a converse for all such phase monitors; for
homogeneous i.i.d. streams with a positive surplus-dephasing exponent it is
supplemented by an explicit achievable block-capacity law.  A
finite-temperature spin-phase bath illustrates the physical separation: hot
probes can suppress system coherences efficiently while losing the local
asymmetry needed to support objective witnesses.

\paragraph{Microscopic QND Hamiltonian and Strong-QD Deficit.---}

We use the standard fragmented-environment architecture of quantum Darwinism, in
which many environmental carriers acquire conditional imprints of a preferred
system observable and are later grouped into independently accessible fragments
\(F\) \cite{OllivierPoulinZurek,OllivierPoulinZurek2005,BlumeKohoutZurek2005,
BlumeKohoutZurek,ZwolakQuanZurek2009,ZwolakQuanZurek2010,RiedelZurek2010}.  In
the repeated-interaction representation \cite{AttalPautrat2006,Ciccarello2022,
LacroixCilluffoHuelgaPlenio2025}, the main text restricts attention to the QND
monitoring sector.  The \(k\)th monitoring event is generated by
\begin{equation}
H_k^{\rm QND}(t)=H_{S,k}^{\Pi}(t)+H_{E_k}(t)+\sum_x\Pi_x\otimes B_{k,x}(t).
\label{eq:Hmain}
\end{equation}
Here \(H_{S,k}^{\Pi}(t)\) is pointer diagonal, \(H_{E_k}(t)\) is the carrier
self-Hamiltonian, and \(B_{k,x}(t)\) conditionally drives carrier \(E_k\) when
the pointer value is \(x\).  
We work in the interaction picture with respect to the pointer-diagonal system
Hamiltonian, or assume that \(H_{S,k}^{\Pi}\) acts as a scalar on each pointer
sector.
The corresponding collision is an ideal controlled
monitor,
\begin{equation}
U_k^0=\sum_x\Pi_x\otimes U_{k,x},\qquad
\rho_{k|x}=U_{k,x}\sigma_kU_{k,x}^\dagger .
\label{eq:condstates-main}
\end{equation}

The environmental input is initially uncorrelated with the system and factorized
across carriers,
\begin{equation}
\rho_{SE}(0)=\rho_S(0)\otimes\bigotimes_{k=1}^N\sigma_k .
\end{equation}
Let \(\{\Pi_x\}\) be the pointer projectors of a finite-dimensional system.  All
entropies and information quantities are measured in bits, with
\(S(\rho)=-\Tr(\rho\log_2\rho)\); logarithms in Chernoff and dephasing
exponents are natural logarithms.  For a joint state \(\rho_{SF}\), define
\begin{equation}
\DeltaPi(\rho_{SF})=\sum_x(\Pi_x\otimes \one_F)\rho_{SF}(\Pi_x\otimes \one_F).
\end{equation}
The pointer probabilities and pointer entropy are
\begin{equation}
p_x=\Tr[(\Pi_x\otimes\one_F)\rho_{SF}],
\qquad
H_\Pi=-\sum_x p_x\log_2 p_x .
\end{equation}
With \(I(A:B)_\omega\) denoting mutual information in the state \(\omega\), the
pointer Holevo information and pointer discord are
\begin{align}
\chi_\Pi(S:F)&=I(S:F)_{\DeltaPi(\rho_{SF})},\\
D_\Pi(S:F)&=I(S:F)_{\rho_{SF}}-\chi_\Pi(S:F).
\end{align}
We use the local strong-QD deficit
\begin{equation}
\sqdd(S:F)\equiv \left[H_\Pi-\chi_\Pi(S:F)\right]_+ +D_\Pi(S:F).
\label{eq:sqddef-main}
\end{equation}
Small \(\sqdd\) means that the fragment stores the pointer value and that its
remaining correlation with the system is classical in the pointer basis.  The
first term is a record-accessibility deficit, while the second is a residual
pointer-discord deficit; the following sections show that they are controlled by
distinct Hamiltonian exponents.


\paragraph{Certified Two-Edge Observer Window for Product Monitoring.---}

The two terms in Eq.~\eqref{eq:sqddef-main} impose complementary
finite-fragment requirements.  The record term asks whether an observer of
\(F\) can distinguish the conditional fragment states associated with
different pointer values; for product monitoring this is a quantum
hypothesis-testing problem, with the quantum Chernoff information setting the
Darwinistic amplification exponent~\cite{AudenaertChernoff,NussbaumSzkola,
ZwolakRiedelZurek2014}.  This gives a lower record-discrimination edge.  The
discord term asks whether the remaining \(S:F\) correlation is classical in
the pointer basis.  Since a mutual-information plateau may still contain
discord~\cite{ZwolakZurek2013,Korbicz2021}, spectrum-broadcast structure,
strong quantum Darwinism, and no-local-broadcasting require the unobserved
complement \(\bar F\) to suppress the residual pointer coherences
~\cite{KorbiczHorodeckiHorodecki2014,HorodeckiKorbiczHorodecki2015,LeOlaya,
PianiHorodeckiHorodecki2008,BrandaoPianiHorodecki2015}.  This gives an upper
surplus-dephasing edge.

We now turn these two requirements into finite-size certificates for exact QND product monitoring. 
For a fragment \(F\subset E\), let \(\barF=E\setminus F\) denote the unobserved
complement.  In exact QND product monitoring,
\begin{equation}
\rho_{F|x}=\bigotimes_{k\in F}\rho_{k|x},
\qquad
\rho_{k|x}=U_{k,x}\sigma_kU_{k,x}^\dagger .
\end{equation}
The two Hamiltonian exponents controlling strong Darwinism are the record exponent
and the surplus-dephasing exponent.

The record exponent of the observed fragment is the multiple quantum Chernoff exponent
\begin{equation}
C_{\rm rec}(F)
=
\min_{x\ne x':\,p_xp_{x'}>0}
\Big[-\log
\inf_{0\le s\le1}
\prod_{k\in F}
\Tr\!\big(
\rho_{k|x}^{s}\rho_{k|x'}^{1-s}
\big) \Big].
\label{eq:chernoff-main}
\end{equation}
It quantifies how well the states \(\{\rho_{F|x}\}\) can be distinguished by an optimal
observer of \(F\).  The minimum is restricted to pointer labels with nonzero prior probability; equivalently, one may assume the finite pointer alphabet has full support.
In a homogeneous i.i.d. stream with \(|F|=m\),
\(C_{\rm rec}(F) \) scales with $m$ linearly.

The residual coherence left by the unobserved complement is
\begin{align}
\Gamma_{\barF}^{xx'}
&=
\prod_{k\in\barF}
\Tr\!\left(U_{k,x}\sigma_kU_{k,x'}^\dagger\right),
\\
R_{\rm dec}(\barF)
&=
\max_{x\ne x'}{}'
\left|\Gamma_{\barF}^{xx'}\right|^2 ,
\label{eq:residual-coherence-main}
\end{align}
where the prime restricts the maximum to active coherences,
\(\Pi_x\rho_S(0)\Pi_{x'}\ne0\).  The surplus-dephasing exponent is
\begin{equation}
C_{\rm dec}(\barF)=-\log R_{\rm dec}(\barF).
\label{eq:cdec-main}
\end{equation}
It quantifies the decohering power still supplied by the environment outside
\(F\).  In a homogeneous i.i.d. stream with \(N\) total carriers
and \(|F|=m\), \(C_{\rm dec}(\barF)\) scales linearly with \(N-m\).

The record exponent therefore sets the lower edge of the observer window: a fragment
that is too small has not accumulated enough distinguishability.  The surplus-dephasing
exponent sets the upper edge: a fragment that is too large leaves too small an unobserved
complement to remove the pointer-basis coherence that still purifies the system.

\begin{theorem}
\label{thm:sharp-window}
\textbf{Certified two-edge window.}
Consider exact QND product monitoring with a finite full-support pointer alphabet.
Under the regular finite-model assumptions, up to additive
\(\mathcal{O}(1)\) constants in the exponent thresholds, the two inequalities
\begin{equation}
\begin{cases}
C_{\rm rec}(F)\ge
\log\frac1\eps+\log\log\frac1\eps+\mathcal O(1),\\[1mm]
C_{\rm dec}(\bar F)\ge
\log\frac1\eps+\mathcal O(1)
\end{cases}
\label{eq:threshold-main}
\end{equation}
are a sufficient finite-accuracy certificate for
\(\sqdd(S:F)=\mathcal O(\eps)\).  The constants may depend on the fixed model and
pointer prior, but not on \(F\) or \(\eps\).  
\end{theorem}


The regular finite-model assumptions and the converse bounds at
the same logarithmic scale are stated in the Supplemental Material (SM).
Theorem~\ref{thm:sharp-window} gives a finite-size certificate which,
counterintuitively, is not monotone in fragment size.  It combines the
record-discrimination requirement for the observed fragment and the
residual-decoherence requirement for its complement as two competing
Hamiltonian thresholds.  The lower edge is an optimal multi-hypothesis
discrimination problem on the observed carriers.  The upper edge is the residual
coherence left by the unobserved carriers.  For a target accuracy \(\eps\),
Eq.~\eqref{eq:threshold-main} certifies objective records only when both edges
are crossed.  A small fragment has not accumulated enough distinguishability,
while a nearly complete fragment captures degrees of freedom that still purify
the system.  This is complementary to recent analyses of
redundancy, consensus, and metrological readout precision, which emphasize that
fragment averages, observer agreement, and measurement choice carry operational
information not contained in a single mutual-information curve
\cite{ChisholmInnocentiPalma2023,ChisholmInnocentiPalma2024,
TouilYanZurek2025,KielyChisholmTouilDeffnerLandiCampbell2026}.

\begin{figure}[t]
\centering
\includegraphics[width=0.8\linewidth]{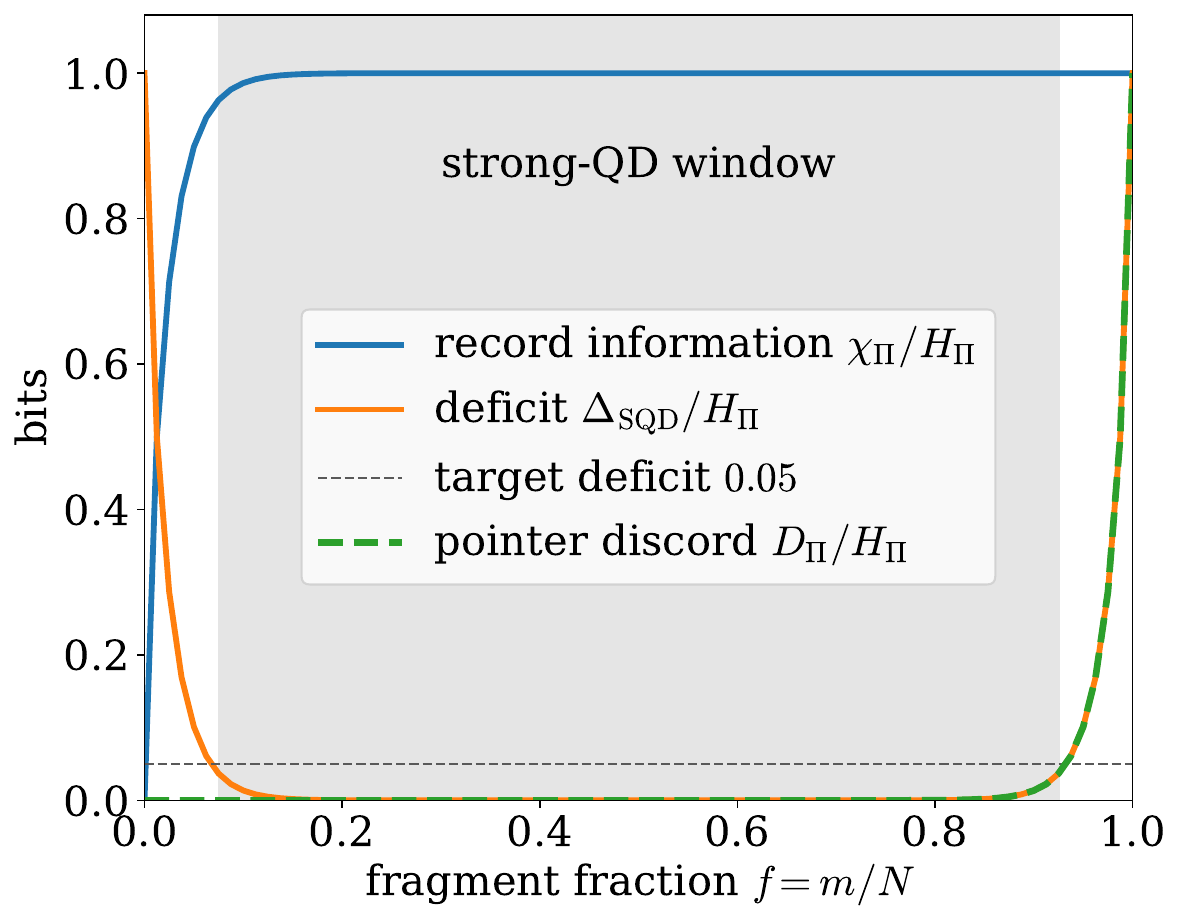}
\caption{Certified two-edge observer window in an i.i.d. binary pure-probe model with equal pointer priors and
single-carrier conditional overlap
\(s=|\langle e_+|e_-\rangle|\in[0,1]\).  The fragment fraction is
\(f=m/N\), with \(m=0,1,\ldots,N\).  The plotted parameters are
\(N=80\), \(s=0.78\), and the shaded observer window is defined by
\(\Delta_{\rm SQD}(m)/H_\Pi<0.05\); since \(H_\Pi=1\) in this binary equal-prior example,
this is \(\Delta_{\rm SQD}(m)<0.05\). }
\label{fig:window-main}
\end{figure}

In a homogeneous i.i.d. stream, these two sufficient exponent inequalities become
especially transparent (Fig. \ref{fig:window-main}).  If each carrier contributes the same record exponent
\(c_{\rm rec}\) and the same surplus-dephasing exponent \(c_{\rm dec}\), then for
\(|F|=m\) and total carrier number \(N\), an \(\epsilon\)-accurate objective
fragment is certified whenever
\begin{equation}
\begin{split}
m&\ge
\frac{\log(1/\epsilon)+\log\log(1/\epsilon)+O(1)}{c_{\rm rec}},\\[1mm]
m&\le
N-\frac{\log(1/\epsilon)+O(1)}{c_{\rm dec}} .
\end{split}
\label{eq:m_window}
\end{equation}
The left boundary is the record edge: if \(F\) is too small, \(m c_{\rm rec}\)
is insufficient to identify the pointer value.  The right boundary is the
surplus-dephasing edge: if \(F\) is too large, then \(\bar F\) is too small and
\((N-m)c_{\rm dec}\) is insufficient to remove the residual pointer-basis
discord between \(S\) and \(F\).  The certified window is nonempty only when the
lower record length plus the required dephasing complement does not exceed \(N\).
Under the additional regularity assumptions used in the supplement, the same two
edges also give converse bounds at logarithmic accuracy.



\paragraph{Asymmetry as the Resource for Redundant Witnesses.---}
\label{sec:asymmetry-resource}

The certified-window theorem identifies the kinematics of objective fragments, but
not the microscopic resource that pays for the record edge.  Here we isolate
that resource in the QND, or phase-monitoring, sector of product monitoring:
when conditional environmental dynamics is generated by a fixed local Abelian
algebra, redundant readable records require symmetry breaking in the initial
environment.  
The basic single-generator case,
\begin{equation}
H_{\rm int}=A_S\otimes G_E,\qquad
A_S=\sum_xa_x\Pi_x,\qquad
G_E=\sum_{k=1}^NG_k,
\end{equation}
is the familiar pure-dephasing Hamiltonian underlying central-spin and
hazy-environment models of redundant records
\cite{BlumeKohoutZurek,ZwolakQuanZurek2009,ZwolakZurek2013,
ZwolakRiedelZurek2014}.  Here \(G_k\) is the Hermitian local generator whose
exponential implements the pointer-conditioned phase rotation of carrier \(k\),
and whose eigenspaces define the charge sectors relative to which asymmetry is
measured.

We allow each carrier to have a finite commuting family of local generators
\begin{equation}
\boldsymbol G_k=(G_k^1,\ldots,G_k^{r_G}),
[G_k^\mu,G_k^\nu]=0,
\end{equation}
with conditional phase action
\begin{equation}
U_{k,x}
=
\exp\!\left[-\ii\sum_{\mu=1}^{r_G}\theta_{k,x}^{\mu}G_k^\mu\right].
\end{equation}
For a fragment \(F\),
\begin{align}
U_{F,x}
&=
\bigotimes_{k\in F}U_{k,x}
\nonumber\\
&=
\exp\!\left[
-\ii\sum_{k\in F}\sum_{\mu=1}^{r_G}
\theta_{k,x}^{\mu}G_k^\mu
\right],
\\
\rho_{F|x}
&=U_{F,x}\sigma_FU_{F,x}^\dagger .
\label{eq:phase-orbit-main}
\end{align}

Let \(\Delta_{\boldsymbol G_F}\) denote the pinching onto the joint eigenspaces
of the local commuting family
\begin{equation}
\boldsymbol G_F=\{G_k^\mu:\ k\in F,\ \mu=1,\ldots,r_G\}.
\end{equation}
The local Abelian asymmetry of \(F\) is
\begin{equation}
A_{\boldsymbol G_F}(\tau)
=
S(\Delta_{\boldsymbol G_F}\tau)-S(\tau).
\label{eq:local-asymmetry-main}
\end{equation}
For a nondegenerate joint charge basis this is the relative entropy of
coherence; with degeneracies it is the relative entropy of asymmetry for the
Abelian algebra \cite{BaumgratzCoherence,MarvianSpekkens}.  The locality is
essential: coherences hidden inside a coarse total charge do not yield
independent records for disjoint observers.
This separates dephasing from witnessing.  If
\(\Delta_{\boldsymbol G_k}(\sigma_k)=\sigma_k\), then
\(U_{k,x}\sigma_kU_{k,x}^\dagger=\sigma_k\) for all \(x\), so carrier \(k\)
stores no readable record.  Yet its single-carrier contribution
\(\Gamma_{xx'}^{(k)}=\Tr(U_{k,x}\sigma_kU_{k,x'}^\dagger)\) to the system
coherence factor may still have modulus below one.  Diagonal charge uncertainty can dephase the
system; readable local records require coherence between charge sectors.

\begin{theorem}
\label{thm:asymmetry-capacity}
Consider an exact product monitor with finite pointer alphabet whose conditional
environmental dynamics is the Abelian phase monitor of
Eq.~\eqref{eq:phase-orbit-main}, and assume that
Eq.~\eqref{eq:local-asymmetry-main} is finite.  Then every fragment \(F\)
obeys the asymmetry converse
\begin{equation}
\chi_\Pi(S:F)
\le
A_{\boldsymbol G_F}(\sigma_F).
\label{eq:asym-bound-main}
\end{equation}
Thus an Abelian phase monitor can convert pre-existing local environmental
asymmetry into pointer information, but cannot create more readable record
information than the fragment's local asymmetry budget.

Consequently, for a product environment \(\sigma_E=\bigotimes_{k=1}^N\sigma_k\), if \(F_1,\ldots,F_R\) are disjoint
\(\delta\)-accurate strong-QD witnesses, meaning
\(\Delta_{\rm SQD}(S:F_i)\le\delta\) for every \(i\), in an \(N\)-carrier environment, then
their maximum number \(R_\delta(N)\) satisfies
\begin{equation}
R_\delta(N) \le 
\frac{\sum_{k=1}^N A_{\boldsymbol G_k}(\sigma_k)}{H_\Pi-\delta},
\qquad 0<\delta<H_\Pi .
\label{eq:red-product-converse-main}
\end{equation}

For a homogeneous i.i.d. stream, \(\sigma_E=\sigma^{\otimes N}\) and
\(U_{k,x}=U_x\), assume that complements of fixed-size blocks have a positive
surplus-dephasing exponent, i.e., for each fixed block size \(m\) there are
constants \(b_m,c_m>0\) such that
\(R_{\rm dec}(\bar F)\le b_m e^{-c_m(N-m)}\) for every \(|F|=m\).
If \(\chi_m\) is the pointer Holevo information
in an \(m\)-carrier block, define
\begin{equation}
\mrec(\delta)
=
\min\{m\ge1:\chi_m\ge H_\Pi-\delta\}.
\label{eq:mrec-main}
\end{equation}
Then, up to the unavoidable integer rounding of block size, the asymptotic
density of independent objective witnesses is
\begin{equation}
\Cred(\delta)
:=
\lim_{N\to\infty}\frac{R_\delta(N)}{N}
=
\frac{1}{\mrec(\delta)}
\label{eq:capacity-main}
\end{equation}
at every continuity point of the integer-valued function \(\mrec(\delta)\).  At
jump points, the corresponding one-sided liminf--limsup statement retains the
same integer block-size rounding.
\end{theorem}

The theorem places local Abelian asymmetry between the Hamiltonian mechanism
and the state-level criteria for objectivity.  Quantum Darwinism introduced
redundancy as the proliferation of environmental records of a pointer
observable \cite{OllivierPoulinZurek,BlumeKohoutZurek,ZurekNatPhys}; spectrum
broadcast structure and strong quantum Darwinism sharpened this into locally
accessible classical information with no residual quantum correlations
\cite{HorodeckiKorbiczHorodecki2015,LeOlaya,Korbicz2021}; generic Darwinism
explains why redundantly accessible information must be classical and tied to
one effective system observable \cite{BrandaoPianiHorodecki2015}.  In the
Abelian phase-monitoring sector, Theorem~\ref{thm:asymmetry-capacity} adds the
resource account: crossing the lower edge of the certified observer window
requires spending local environmental asymmetry.


\paragraph{Example: Tilted Thermal Spin-Phase Bath.---}
\label{sec:spinphase-example}

\begin{figure}[t]
\centering
\includegraphics[width=0.92\columnwidth]{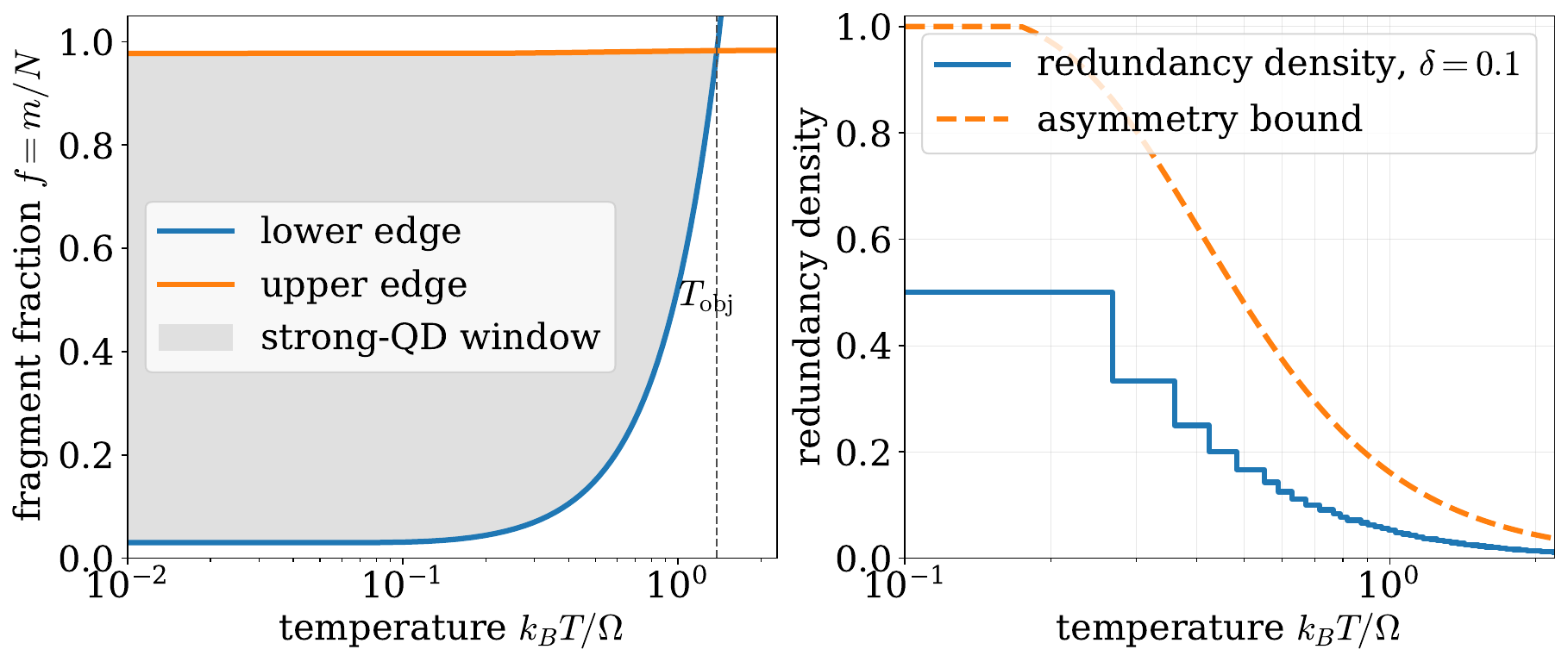}
\caption{The same spin-phase bath illustrates both the certified two-edge window and the
asymmetry-capacity law for equal pointer priors.  Parameters are the carrier Bloch tilt
\(\vartheta=0.40\pi\), phase kick \(\phi=0.40\pi\), carrier number \(N=120\), window target
\(\epsilon=10^{-2}\), and record-deficit tolerance \(\delta=0.10\).  The temperature axis is
the dimensionless ratio \(k_BT/\Omega\).  Left: the lower edge uses the plotted threshold
\(\log(1/\epsilon)+\log\log(1/\epsilon)=6.13\), while the upper edge uses
\(\log(1/\epsilon)=4.61\), with the unspecified \(\mathcal O(1)\) constants
set to zero.  The shaded region is the strong-QD window.  Right: the exact block capacity
\(1/m_{\rm rec}(\delta;T)\), where \(m_{\rm rec}\) is the smallest block size reaching the
tolerance \(\delta\), is computed by the Schur-block method in Eq.~\eqref{eq:app-spinphase-schur},
searching block sizes \(1\le m\le110\).  It falls in integer steps and always lies below the
asymmetry converse \(A_G(\sigma_T)/(1-\delta)\), where \(\sigma_T\) is the one-carrier thermal
state.}
\label{fig:spin_bath}
\end{figure}

The separation between decoherence and objective witnessing is already visible in a minimal spin bath.  Let a qubit pointer with equal pointer priors be monitored by independent spin-$1/2$ carriers.  The Pauli operators on carrier \(k\) are denoted by \(\tau_\mu^{(k)}\), and \(\one\) denotes the carrier identity.  Before the collision, carrier $k$ is prepared in
\begin{equation}
\sigma_k=\frac12\!\left[\one+r_k
\left(\sin\vartheta_k\,\tau_x^{(k)}
+\cos\vartheta_k\,\tau_z^{(k)}\right)\right],
\label{eq:tilted-thermal-state}
\end{equation}
where $r_k=\tanh(\beta\Omega_k/2)$, \(\beta=1/(k_BT)\), \(\Omega_k\) is the carrier energy scale, and the monitoring interaction is the QND phase kick
\begin{equation}
H_{{\rm int},k}=\frac{g_k}{2}\sigma_z^S\otimes\tau_z^{(k)},
\qquad
U_{k,\pm}=\exp\!\left(\mp\frac{i\phi_k}{2}\tau_z^{(k)}\right)
\label{eq:spin-phase-collision}
\end{equation}
Here $\phi_k=g_kt$, \(g_k\) is the coupling strength, and \(t\) is the collision duration.  Thus $\rho_{k|\pm}=U_{k,\pm}\sigma_k U_{k,\pm}^\dagger$ lies on an Abelian phase orbit generated by
$G_k=\tau_z^{(k)}/2$.  The model is therefore simultaneously an exact product monitor, to which the two-edge window applies, and an Abelian phase monitor, to which the asymmetry bound applies.  We write
\(h_2(q)=-q\log_2q-(1-q)\log_2(1-q)\) for the binary entropy.

For one carrier, the three relevant quantities are (see the SM for derivation)
\begin{align}
c_{{\rm dec},k}
&=-\log\!\left(\cos^2\phi_k+r_k^2\cos^2\vartheta_k\sin^2\phi_k\right),
\label{eq:spin-cdec}
\\
c_{{\rm rec},k}
&=-\log\!\left[1-\left(1-\sqrt{1-r_k^2}\right)
\sin^2\vartheta_k\sin^2\phi_k\right],
\label{eq:spin-crec}
\\
A_k
&=A_{G_k}(\sigma_k)
=h_2\!\left(\frac{1+r_k\cos\vartheta_k}{2}\right)
-h_2\!\left(\frac{1+r_k}{2}\right).
\label{eq:spin-asymmetry}
\end{align}
The unobserved carriers suppress residual pointer coherence through $c_{\rm dec}$, whereas the observed carriers become readable witnesses only through the transverse phase asymmetry measured by $A_k$.

At high temperature, $k_BT\gg\Omega_k$ and $r_k\simeq\Omega_k/(2k_BT)$.  For fixed $\vartheta_k$ and $\phi_k$,
\begin{align}
c_{{\rm rec},k}&\simeq
\frac{r_k^2}{2}\sin^2\vartheta_k\sin^2\phi_k,
\\
A_k&\simeq
\frac{r_k^2\sin^2\vartheta_k}{2\ln 2},
\\
c_{{\rm dec},k}&\to -\log\cos^2\phi_k .
\end{align}
Thus heating suppresses the record exponent and the asymmetry budget as $T^{-2}$, while the dephasing exponent can remain finite.  In the charge-diagonal limit $\vartheta_k=0$, one has $A_k=c_{{\rm rec},k}=0$ exactly, although $c_{{\rm dec},k}$ can be nonzero.  This is dephasing without witnessing.

For a homogeneous bath, $r_k=r$, $\vartheta_k=\vartheta$, and $\phi_k=\phi$; we write \(\sigma_T\) for its one-carrier state at temperature \(T\).  A fragment of $m$ carriers in an $N$-carrier environment is certified by Eq.~\eqref{eq:m_window}.
The lower edge therefore moves to larger fragments as temperature increases, while the upper edge remains controlled by the residual dephasing power of the unobserved complement.  Define the block record length
\begin{equation}
 m_{\rm rec}(\delta;T)=\min\{m:\chi_m(T)\ge1-\delta\},
\label{eq:spin-mrec-main}
\end{equation}
where $H_\Pi=1$ because the pointer priors are equal.  The same model also illustrates the capacity statement: at continuity points of $m_{\rm rec}$, the asymptotic redundancy density is
\begin{equation}
C_{\rm red}(\delta;T) \equiv \lim \limits_{N\to \infty} \frac{R_\delta(N)}{N}
=\frac{1}{m_{\rm rec}(\delta;T)}
\le \frac{A_G(\sigma_T)}{1-\delta}
\label{eq:spin-capacity}
\end{equation}
up to integer block-size rounding.  The critical temperature $T_\text{obj}$ for a fixed $N$ and $\epsilon$ is defined by equality of the two edges in Eq.~\eqref{eq:m_window}; above it the required record block plus the required dephasing complement exceeds the available environment.  Figure~\ref{fig:spin_bath} shows the two consequences in the same microscopic model: the certified strong-QD window closes as the record edge rises, and the redundancy density falls in discrete block-size steps as thermal mixing removes local phase asymmetry.


\paragraph{Discussion and outlook.---}
This Letter gives finite-fragment Hamiltonian certificates for objective
records.  In exact QND product monitoring, strong quantum Darwinism requires
two simultaneous conditions: the observed fragment must resolve the pointer
alternatives, and the unobserved complement must still erase the residual
pointer coherence.  Objectivity is therefore windowed rather than monotone in
fragment size, refining the usual redundancy picture of quantum Darwinism and
its strong/SBS variants
\cite{OllivierPoulinZurek,ZurekNatPhys,HorodeckiKorbiczHorodecki2015,
LeOlaya,Korbicz2021}.

For Abelian phase monitors, the lower edge of this window has a resource
meaning.  A carrier can become a local witness only by spending asymmetry
relative to the coupling generator; dephasing alone is not enough
\cite{BaumgratzCoherence,MarvianSpekkens}.  The tilted spin-phase bath makes
the separation explicit: thermal mixing reduces transverse asymmetry, raises
the record edge, and lowers the block redundancy density even when the residual
dephasing exponent remains finite
\cite{ZwolakQuanZurek2009,ZwolakQuanZurek2010,ZwolakRiedelZurek2014}.

The next step is to test these two edges in platforms where both system
coherence and fragment information can be measured, such as central-spin
systems, ancilla arrays, superconducting readout chains, and engineered
collision models
\cite{TouilYanGirolamiDeffnerZurek2022,ZhuEtAlSciAdv2025,
Ciccarello2022,DoucetDeffner2024}.  Open problems include non-Abelian
monitoring, correlated inputs, interacting baths, and finite-time departures
from product QND dynamics.

\bibliography{ref}

\clearpage

\onecolumngrid

\setcounter{secnumdepth}{2}
\renewcommand{\thesection}{S\arabic{section}.  }
\renewcommand{\thesubsection}{S\arabic{section}.\arabic{subsection}}

\makeatletter
\def\@sectioncntformat#1{\csname the#1\endcsname}
\makeatother

\begin{center}
\textbf{\large Supplemental Materials: Hamiltonian Thresholds for Objective Records}

\vspace{1em}

{Jie Gu}

{Chengdu Academy of Education Sciences, Chengdu 610036, China}
 
\end{center}

\setcounter{section}{0}
\setcounter{equation}{0}
\setcounter{figure}{0}
\setcounter{table}{0}
\renewcommand{\thepage}{S\arabic{page}}
\setcounter{page}{1}
\renewcommand{\theequation}{S\arabic{equation}}
\renewcommand{\thefigure}{S\arabic{figure}}
\renewcommand{\thetable}{S\arabic{table}}
\makeatletter
\renewcommand{\theHsection}{S\arabic{section}}
\renewcommand{\theHsubsection}{S\arabic{section}.\arabic{subsection}}
\renewcommand{\theHequation}{S\arabic{equation}}
\renewcommand{\theHfigure}{S\arabic{figure}}
\renewcommand{\theHtable}{S\arabic{table}}
\makeatother
\renewcommand{\bibnumfmt}[1]{[S#1]}
\renewcommand{\citenumfont}[1]{S#1}

\section{QND controlled product monitoring and finite-size estimates}
\label{app:finite-size}

This appendix proves the finite-size estimates used in the certified two-edge theorem,
Theorem~\ref{thm:sharp-window}, and explains how the QND microscopic Hamiltonian
in Eq.~\eqref{eq:Hmain} leads to the exact product-monitoring dynamics in
Eq.~\eqref{eq:condstates-main}.  The final subsection extends this QND
construction to weak non-QND perturbations and records the associated stability
correction.

We start from an initially uncorrelated environment,
\begin{equation}
\rho_{SE}(0)=\rho_S(0)\otimes\bigotimes_{k=1}^{N}\sigma_k .
\label{eq:app-initial-product}
\end{equation}
Since the QND Hamiltonian in Eq.~\eqref{eq:Hmain} is diagonal in the pointer
projectors, the propagator for the \(k\)th collision has the controlled form
\begin{equation}
U_k^0=\sum_x \Pi_x\otimes U_{k,x}.
\label{eq:app-controlled-unitary}
\end{equation}
The conditional carrier state is therefore
\begin{equation}
\rho_{k|x}=U_{k,x}\sigma_k U_{k,x}^\dagger,
\label{eq:app-conditional-carrier}
\end{equation}
and, for a product monitor, the conditional state of a fragment \(F\) factorizes:
\begin{equation}
\rho_{F|x}=\bigotimes_{k\in F}\rho_{k|x}.
\label{eq:app-conditional-fragment}
\end{equation}

Tracing out the complement \(\barF=E\setminus F\) gives the controlled reference
state of \(S:F\),
\begin{align}
\rhozero_{SF}
&=\sum_{x,x'}
\Pi_x\rho_S(0)\Pi_{x'}
\otimes
\left[
\bigotimes_{k\in F}
U_{k,x}\sigma_k U_{k,x'}^\dagger
\right]
\Gamma_{\barF}^{xx'},
\label{eq:app-rhoSF-controlled}
\end{align}
where
\begin{equation}
\Gamma_{\barF}^{xx'}
=
\prod_{k\in\barF}
\Tr\!\left(U_{k,x}\sigma_kU_{k,x'}^\dagger\right).
\label{eq:app-Gamma}
\end{equation}
For \(x=x'\), \(\Gamma_{\barF}^{xx}=1\).  For \(x\ne x'\),
\(\Gamma_{\barF}^{xx'}\) is the residual pointer coherence left after the
unobserved complement has been discarded.  Pointer dephasing removes precisely
these off-diagonal blocks:
\begin{equation}
\DeltaPi(\rhozero_{SF})
=
\sum_x
\Pi_x\rho_S(0)\Pi_x
\otimes
\bigotimes_{k\in F}\rho_{k|x}.
\label{eq:app-dephased-controlled}
\end{equation}
Thus the dephased state is a classical-quantum ensemble with label \(x\) and
record state \(\rho_{F|x}\).

\subsection{Record contribution}

The record part of the strong-QD deficit is
\begin{equation}
[H_\Pi-\chi_\Pi(S:F)]_+ .
\label{eq:app-record-deficit}
\end{equation}
For the controlled state in Eq.~\eqref{eq:app-dephased-controlled},
\(\chi_\Pi(S:F)\) is the Holevo information of the ensemble
\begin{equation}
\mathcal E_F=\{p_x,\rho_{F|x}\}.
\label{eq:app-ensemble}
\end{equation}
Let \(P_e^\star(F)\) be the minimum probability of incorrectly guessing \(x\) by
an arbitrary POVM on \(F\).  For any POVM with classical outcome \(Y\), the
Holevo bound gives
\begin{equation}
I(X:Y)\le \chi_\Pi(S:F)_{\rhozero}.
\label{eq:app-holevo-bound}
\end{equation}
Using \(I(X:Y)=H(X)-H(X|Y)\) and \(H(X)=H_\Pi\), this implies
\begin{equation}
H_\Pi-\chi_\Pi(S:F)_{\rhozero}
\le H(X|Y).
\label{eq:app-record-to-classical-conditional}
\end{equation}
Choose a POVM attaining \(P_e^\star(F)\), up to an arbitrarily small slack.  Fano's
inequality for a \(q\)-valued pointer alphabet gives
\begin{equation}
H(X|Y)
\le
h_2(P_e^\star)+P_e^\star\log_2(q-1),
\label{eq:app-fano}
\end{equation}
where \(h_2\) is the binary entropy.  Hence
\begin{equation}
H_\Pi-\chi_\Pi(S:F)_{\rhozero}
\le
h_2(P_e^\star)+P_e^\star\log_2(q-1).
\label{eq:app-record-upper}
\end{equation}
This is the finite-size direction used for the lower edge of
Theorem~\ref{thm:sharp-window}: a fragment that can discriminate the pointer
label with small error has small Holevo deficit.  For a binary pointer,
\(q=2\), the second term vanishes.

\subsection{Surplus-dephasing contribution}

The pointer-basis discord is
\begin{equation}
D_\Pi(S:F)=I(S:F)_\rho-I(S:F)_{\DeltaPi(\rho)}.
\label{eq:app-discord-def}
\end{equation}
For the controlled reference state, the only difference between \(\rhozero_{SF}\)
and \(\DeltaPi(\rhozero_{SF})\) is the off-diagonal operator
\begin{equation}
X_{SF}=\rhozero_{SF}-\DeltaPi(\rhozero_{SF}),
\label{eq:app-X-def}
\end{equation}
namely
\begin{align}
X_{SF}
&=
\sum_{x\ne x'}
\Pi_x\rho_S(0)\Pi_{x'}
\otimes
\left[
\bigotimes_{k\in F}
U_{k,x}\sigma_kU_{k,x'}^\dagger
\right]
\Gamma_{\barF}^{xx'} .
\label{eq:app-X-blocks}
\end{align}
Each environmental factor has trace norm one,
\begin{equation}
\left\|U_{k,x}\sigma_kU_{k,x'}^\dagger\right\|_1=
\|\sigma_k\|_1=1.
\label{eq:app-env-trace-norm}
\end{equation}
Therefore
\begin{align}
\frac12\left\|\rhozero_{SF}-\DeltaPi(\rhozero_{SF})\right\|_1
&\le
\frac12
\sum_{x\ne x'}
\left\|\Pi_x\rho_S(0)\Pi_{x'}\right\|_1
\left|\Gamma_{\barF}^{xx'}\right|
\nonumber\\
&\le
C_\rho
\max_{x\ne x'}{}'\left|\Gamma_{\barF}^{xx'}\right|,
\label{eq:app-coherence-trace-bound}
\end{align}
where the prime restricts the maximum to active coherences and
\begin{equation}
C_\rho=
\frac12
\sum_{x\ne x'}
\left\|\Pi_x\rho_S(0)\Pi_{x'}\right\|_1 .
\label{eq:app-Crho}
\end{equation}
Let
\begin{equation}
\eta_0=
\frac12\left\|\rhozero_{SF}-\DeltaPi(\rhozero_{SF})\right\|_1 .
\label{eq:app-eta0}
\end{equation}
Since \(\rhozero_F=\DeltaPi(\rhozero_{SF})_F\), the discord can be written as a
difference of conditional entropies and system entropies.  The
Alicki-Fannes-Winter continuity bound for conditional entropy, together with the
ordinary Fannes-Audenaert bound for the system entropy, gives for
\(0\le\eta_0\le1/2\)
\begin{equation}
D_\Pi(S:F)_{\rhozero}
\le
g_{d_S}(\eta_0),
\label{eq:app-discord-continuity}
\end{equation}
where \(d_S=\dim\cH_S\), with the nonoptimized envelope
\begin{equation}
g_d(\eta)=4\eta\log_2 d+4h_2(\eta).
\label{eq:app-g-envelope}
\end{equation}
The relevant dimension is \(d_S\), not the fragment dimension, because the
conditional-entropy continuity bound depends on the conditioned subsystem.
Combining Eqs.~\eqref{eq:app-coherence-trace-bound} and
\eqref{eq:app-discord-continuity} yields
\begin{equation}
D_\Pi(S:F)_{\rhozero}
\le
 g_{d_S}\!\left(
C_\rho
\max_{x\ne x'}{}'|\Gamma_{\barF}^{xx'}|
\right).
\label{eq:app-discord-upper}
\end{equation}
This is the finite-size surplus-dephasing estimate used for the upper edge of
Theorem~\ref{thm:sharp-window}.

\subsection{Stability under non-QND terms}

The main text treats the QND Hamiltonian in Eq.~\eqref{eq:Hmain}.  To include
weak departures from QND monitoring, replace it by
\begin{equation}
H_k^{\rm act}(t)=H_k^{\rm QND}(t)+V_k^\perp(t),
\label{eq:app-H-nonqnd}
\end{equation}
where \(V_k^\perp(t)\) contains the terms that connect different pointer
sectors during the readout.  Define the accumulated non-QND action
\begin{equation}
\xi_N=\sum_{k=1}^N\int_0^{\tau_k}\|V_k^\perp(t)\|\,\dd t ,
\label{eq:app-xiN}
\end{equation}
with \(\tau_k\) the duration of the \(k\)th collision.  Let \(U^0\) be the
product controlled propagator generated by the QND part and \(U^{\rm act}\) the
propagator generated by Eq.~\eqref{eq:app-H-nonqnd}.  In the interaction picture
with respect to the controlled part,
\begin{equation}
W=(U^0)^\dagger U^{\rm act}
=
\mathcal T
\exp\!\left[
-\ii\sum_{k=1}^{N}\int_0^{\tau_k}V_{k,I}^\perp(t)\,\dd t
\right],
\label{eq:app-W}
\end{equation}
where unitary conjugation gives
\(\|V_{k,I}^\perp(t)\|=\|V_k^\perp(t)\|\).  The Dyson expansion gives
\begin{equation}
\|W-\one\|
\le
\exp(\xi_N)-1
=
\xi_N+\cO(\xi_N^2).
\label{eq:app-W-bound}
\end{equation}
Hence, for every density operator,
\begin{align}
\frac12\|W\rho W^\dagger-\rho\|_1
&\le \|W-\one\|.
\label{eq:app-state-stability-step}
\end{align}
After applying the controlled unitary and tracing out arbitrary environmental
subsystems,
\begin{equation}
\frac12\left\|\rhoact_{SF}-\rhozero_{SF}\right\|_1
\le
\xi_N+\cO(\xi_N^2).
\label{eq:app-state-stability}
\end{equation}
Thus the exact product-monitor estimates above remain valid for the full
Hamiltonian, with the two phase boundaries broadened only by continuity
corrections controlled by \(\xi_N\).

\section{Certified observer window and regular-model converse bounds}
\label{app:sharp-window}

This appendix proves the sufficient direction of Theorem~\ref{thm:sharp-window}
and records the corresponding regular-model converse bounds.  The proof separates
the record edge, controlled by optimal discrimination of the observed fragment,
from the residual-coherence edge, controlled by the unobserved complement.

\subsection{Record edge}

The record edge concerns the first term in Eq.~\eqref{eq:sqddef-main}.  The finite-size estimates above give the direct implication
\begin{equation}
P_e^\star(F)\to0
\quad\Longrightarrow\quad
H_\Pi-\chi_\Pi(S:F)\to0,
\label{eq:app-record-direct}
\end{equation}
because the right-hand side of Eq.~\eqref{eq:app-record-upper} tends to zero with
\(P_e^\star(F)\).

For the regular-model converse we use the record-regularity assumption.
For the finite-alphabet product ensembles considered here, record regularity
means that vanishing Holevo deficit and vanishing optimal discrimination error
are equivalent.  Equivalently, there are functions \(r_+(\eta)\) and
\(r_-(\eta)\), both tending to zero with \(\eta\), such that
\begin{align}
P_e^\star(F)\le\eta
&\quad\Longrightarrow\quad
H_\Pi-\chi_\Pi(S:F)\le r_+(\eta),
\label{eq:app-record-regular-plus}
\\
H_\Pi-\chi_\Pi(S:F)\le\eta
&\quad\Longrightarrow\quad
P_e^\star(F)\le r_-(\eta).
\label{eq:app-record-regular-minus}
\end{align}
The first implication is supplied by Eq.~\eqref{eq:app-record-upper}.  The second
excludes pathological product ensembles in which the Holevo information approaches
\(H_\Pi\) while the pointer labels remain operationally indistinguishable.  Under
this assumption,
\begin{equation}
H_\Pi-\chi_\Pi(S:F)\to0
\quad\Longleftrightarrow\quad
P_e^\star(F)\to0.
\label{eq:app-record-iff}
\end{equation}

Let
\begin{equation}
K_{\rm rec}(F)=-\log P_e^\star(F).
\label{eq:app-Krec}
\end{equation}
For a product fragment,
\begin{equation}
\rho_{F|x}=\bigotimes_{k\in F}\rho_{k|x}.
\label{eq:app-product-record-states}
\end{equation}
Thus the pairwise Chernoff quantity factorizes:
\begin{align}
\inf_{0\le s\le1}
\Tr\!\left[\rho_{F|x}^{s}\rho_{F|x'}^{1-s}\right]
&=
\inf_{0\le s\le1}
\prod_{k\in F}
\Tr\!\left(\rho_{k|x}^{s}\rho_{k|x'}^{1-s}\right).
\label{eq:app-chernoff-factorization}
\end{align}
The multiple quantum Chernoff theorem, together with the Chernoff-regularity
assumption used for the regular-model converse, identifies the optimal multi-hypothesis error
exponent with the worst pairwise exponent up to subexponential prefactors:
\begin{equation}
K_{\rm rec}(F)
=
C_{\rm rec}(F)+o(C_{\rm rec}(F)),
\label{eq:app-multiple-chernoff}
\end{equation}
where \(C_{\rm rec}(F)\) is Eq.~\eqref{eq:chernoff-main}.  In particular,
\begin{equation}
K_{\rm rec}(F)\to\infty
\quad\Longleftrightarrow\quad
C_{\rm rec}(F)\to\infty.
\label{eq:app-K-Crec}
\end{equation}
Equation~\eqref{eq:app-multiple-chernoff} fixes only the leading exponential
rate and, by itself, does not control finite-size subexponential prefactors.
For the finite-accuracy implication we therefore use the dimension-free
one-shot pairwise bound for multiple quantum hypothesis testing
\cite{ChengLiu2026}.  Writing
\begin{equation}
C_{xx'}(F)
=
-\log\inf_{0\le s\le1}
\Tr\!\left[\rho_{F|x}^{s}\rho_{F|x'}^{1-s}\right],
\qquad
C_{\rm rec}(F)=\min_{x\ne x'}C_{xx'}(F),
\label{eq:app-pairwise-Chernoff}
\end{equation}
the one-shot bound gives, for a full-support alphabet of fixed size \(q\),
\begin{align}
P_e^\star(F)
&\le
4\sum_{x<x'}\exp[-C_{xx'}(F)]
\nonumber\\
&\le
2q(q-1)\exp[-C_{\rm rec}(F)].
\label{eq:app-one-shot-multiple}
\end{align}
Consequently,
\begin{equation}
K_{\rm rec}(F)
\ge
C_{\rm rec}(F)-\log[2q(q-1)].
\label{eq:app-K-Crec-finite}
\end{equation}
Unlike Eq.~\eqref{eq:app-multiple-chernoff}, this is a nonasymptotic estimate
with a fragment-independent additive constant.

At finite accuracy \(\eps\), set
\begin{equation}
\Phi_q(p)=h_2(p)+p\log_2(q-1),
\label{eq:app-Fano-envelope}
\end{equation}
and let \(\Phi_q^{-1}\) denote the inverse of \(\Phi_q\) on its small-error
branch.  Equation~\eqref{eq:app-record-upper} gives the exact discrimination
threshold
\begin{equation}
K_{\rm rec}(F)\ge \log\frac{1}{\Phi_q^{-1}(\eps)}
\quad\Longrightarrow\quad
H_\Pi-\chi_\Pi(S:F)\le\eps .
\label{eq:app-record-K-threshold}
\end{equation}
Combining Eqs.~\eqref{eq:app-K-Crec-finite} and
\eqref{eq:app-record-K-threshold} gives the explicit sufficient Chernoff
threshold
\begin{equation}
C_{\rm rec}(F)\ge
\log\frac{2q(q-1)}{\Phi_q^{-1}(\eps)}
=
\log\frac{1}{\eps}+\log\log\frac{1}{\eps}+\cO(1)
\quad\Longrightarrow\quad
H_\Pi-\chi_\Pi(S:F)\le\eps ,
\label{eq:app-record-threshold}
\end{equation}
where we used
\(\Phi_q^{-1}(\eps)=\Theta\!\left(\eps/\log(1/\eps)\right)\) as
\(\eps\downarrow0\).  Thus the \(\log\log(1/\eps)\) term comes from the
small-error inversion of the Fano envelope, while the finite-size
multi-hypothesis discrimination prefactor contributes only the displayed
\(\log[2q(q-1)]\), which is absorbed into \(\cO(1)\).
For the regular-model converse, if \(\sqdd(S:F)\le\eps\), then the record term alone gives
\(H_\Pi-\chi_\Pi(S:F)\le\eps\), hence record regularity forces
\(P_e^\star(F)\le r_-(\eps)\).  Therefore
\begin{equation}
C_{\rm rec}(F)+o(C_{\rm rec}(F))
\ge
\log\frac{1}{r_-(\eps)} .
\label{eq:app-record-threshold-converse}
\end{equation}
If the record exponent remains below this scale, no measurement on \(F\) can
supply the required pointer record.

\subsection{Residual-coherence edge}

The upper edge concerns the discord term in Eq.~\eqref{eq:sqddef-main}.  Define
\begin{equation}
\lambda_{\barF}
=
\max_{x\ne x'}{}'|\Gamma_{\barF}^{xx'}|,
\qquad
R_{\rm dec}(\barF)=\lambda_{\barF}^2,
\label{eq:app-lambda-Rdec}
\end{equation}
where the prime again restricts to active coherences.  Equation
\eqref{eq:app-discord-upper} gives
\begin{equation}
D_\Pi(S:F)_{\rhozero}
\le
 g_{d_S}\!\left(C_\rho\lambda_{\barF}\right)
=
 g_{d_S}\!\left(C_\rho\sqrt{R_{\rm dec}(\barF)}\right).
\label{eq:app-D-upper-R}
\end{equation}
Thus residual coherence tending to zero is sufficient for vanishing
pointer-basis discord:
\begin{equation}
R_{\rm dec}(\barF)\to0
\quad\Longrightarrow\quad
D_\Pi(S:F)_{\rhozero}\to0.
\label{eq:app-dec-direct}
\end{equation}

For the converse, set
\begin{equation}
\omega_{SF}=\DeltaPi(\rhozero_{SF}),
\qquad
X_{SF}=\rhozero_{SF}-\omega_{SF}.
\label{eq:app-omega-X}
\end{equation}
The additional nondegenerate active-coherence assumption used for the converse
means that the active diagonal blocks of \(\omega_{SF}\) are bounded away from the
singular directions on which \(X_{SF}\) has support.  The entropy can therefore
be expanded around \(\omega_{SF}\).  For a full-rank state \(\omega\) and a
traceless perturbation \(X\),
\begin{equation}
S(\omega+X)
=
S(\omega)
-\Tr(X\log_2\omega)
-
\frac{1}{2\ln2}\langle X,\Omega_\omega^{-1}(X)\rangle
+
\cO(\|X\|_1^3),
\label{eq:app-entropy-expansion}
\end{equation}
where \(\Omega_\omega^{-1}\) is the positive BKM inverse.  Here the linear terms
vanish because \(X_{SF}\) and \(X_S=\Tr_FX_{SF}\) are off-diagonal in pointer
blocks, whereas \(\log\omega_{SF}\) and \(\log\omega_S\) are block diagonal:
\begin{equation}
\Tr(X_{SF}\log\omega_{SF})=
\Tr(X_S\log\omega_S)=0.
\label{eq:app-linear-zero}
\end{equation}
Using \(\rhozero_F=\omega_F\), the discord is
\begin{equation}
D_\Pi(S:F)
=
[S(\rhozero_S)-S(\omega_S)]
-
[S(\rhozero_{SF})-S(\omega_{SF})].
\label{eq:app-D-entropy-difference}
\end{equation}
Substituting Eq.~\eqref{eq:app-entropy-expansion} gives
\begin{align}
D_\Pi(S:F)_{\rhozero}
&=
\frac{1}{2\ln2}
\left[
\langle X_{SF},\Omega_{\omega_{SF}}^{-1}(X_{SF})\rangle
-
\langle X_S,\Omega_{\omega_S}^{-1}(X_S)\rangle
\right]
+
\cO(\|X_{SF}\|_1^3).
\label{eq:app-D-quadratic}
\end{align}
The square bracket is nonnegative by monotonicity of the BKM metric under partial
trace.  The nondegeneracy condition makes it strictly positive on the active
coherence directions.  Hence, for sufficiently small residual coherence, there
exist constants \(c_- ,c_+>0\), independent of the fragment size, such that
\begin{equation}
c_-R_{\rm dec}(\barF)
\le
D_\Pi(S:F)_{\rhozero}
\le
c_+R_{\rm dec}(\barF)+\cO\!\left(R_{\rm dec}(\barF)^{3/2}\right).
\label{eq:app-D-two-sided-regular}
\end{equation}
Without full-rank regularity the upper side may be replaced by the continuity
bound in Eq.~\eqref{eq:app-D-upper-R}; the lower side is the converse needed for
the upper edge.  If \(D_\Pi(S:F)\le\eps\), then
\(R_{\rm dec}(\barF)\le\eps/c_-\), so
\begin{equation}
C_{\rm dec}(\barF):=-\log R_{\rm dec}(\barF)
\ge
\log\frac{c_-}{\eps} .
\label{eq:app-dec-threshold-converse}
\end{equation}
Conversely, the upper side of Eq.~\eqref{eq:app-D-two-sided-regular} gives a
matching sufficient logarithmic scale.  For sufficiently small \(\eps\), there is
a fragment-independent constant \(c_+'>0\) such that
\begin{equation}
C_{\rm dec}(\barF)\ge
\log\frac{c_+'}{\eps}
=
\log\frac{1}{\eps}+\log c_+'
\quad\Longrightarrow\quad
D_\Pi(S:F)\le\eps .
\label{eq:app-dec-threshold}
\end{equation}
Without the nondegenerate quadratic expansion, the continuity estimate
Eq.~\eqref{eq:app-D-upper-R} gives the weaker but still logarithmic sufficient
condition
\begin{equation}
C_{\rm dec}(\barF)
\ge
2\log\frac{C_\rho}{g_{d_S}^{-1}(\eps)}.
\label{eq:app-dec-continuity-scale}
\end{equation}

Combining the record threshold in Eq.~\eqref{eq:app-record-threshold} with the
surplus-dephasing threshold in Eq.~\eqref{eq:app-dec-threshold} proves the
sufficiency direction of Theorem~\ref{thm:sharp-window}.  Under the additional
regularity assumptions used above, the two converse arguments show that an
\(\eps\)-accurate strong-QD witness lies inside the same two-edge window up to
the finite-alphabet, Chernoff-regularity, and active-coherence constants displayed
above.

\subsection{Exactly solvable finite model used for Fig.~1}
\label{sm:fig1-model}

This section specifies the exactly solvable finite model used to generate
Fig.~1 of the main text.  The purpose of the figure is only to visualize the
two competing requirements for a certified strong-QD witness: the observed
fragment has to carry an accessible pointer record, while the unobserved
complement has to decohere the pointer.

We consider a binary pointer with equal priors,
\begin{equation}
p_+=p_-=\frac12,\qquad H_\Pi=1 ,
\end{equation}
and an environment of \(N\) identical pure probes.  The global branching state is
\begin{equation}
|\Psi\rangle_{S E}
=
\frac{1}{\sqrt2}
\left(
|+\rangle_S |e_+\rangle^{\otimes N}
+
|-\rangle_S |e_-\rangle^{\otimes N}
\right),
\end{equation}
where the single-probe conditional overlap is
\begin{equation}
s = |\langle e_+|e_-\rangle|,\qquad 0<s<1 .
\end{equation}
For a fragment \(F\) containing \(m\) probes, the conditional fragment states are
\begin{equation}
\rho_{F|\pm}
=
|e_\pm\rangle\langle e_\pm|^{\otimes m},
\end{equation}
so their overlap is \(s^m\).  The unobserved complement \(\bar F\) contains
\(N-m\) probes and has conditional overlap \(s^{N-m}\).

Throughout this subsection entropies are measured in bits, and
\begin{equation}
h_2(q)=-q\log_2 q-(1-q)\log_2(1-q)
\end{equation}
denotes the binary entropy.  The mixture of two equiprobable pure states with
overlap \(r\) has nonzero eigenvalues
\begin{equation}
\lambda_\pm(r)=\frac{1\pm r}{2}.
\end{equation}
Therefore the Holevo record in a fragment of size \(m\) is
\begin{equation}
\chi_\Pi(m)
=
S\!\left[
\frac12 \rho_{F|+}+\frac12 \rho_{F|-}
\right]
=
h_2\!\left(\frac{1+s^m}{2}\right).
\end{equation}

The reduced entropy of the pointer is determined by the full-environment
overlap \(s^N\),
\begin{equation}
S(\rho_S)=h_2\!\left(\frac{1+s^N}{2}\right),
\end{equation}
while the entropy of \(SF\) is, by purity of the global state, the entropy of
the complement \(\bar F\),
\begin{equation}
S(\rho_{SF})=S(\rho_{\bar F})
=
h_2\!\left(\frac{1+s^{N-m}}{2}\right).
\end{equation}
The pointer-basis discord contribution used in the main text is therefore
\begin{equation}
D_\Pi(m)
=
I(S:F)-\chi_\Pi(S:F)
=
S(\rho_S)-S(\rho_{SF})
=
h_2\!\left(\frac{1+s^N}{2}\right)
-
h_2\!\left(\frac{1+s^{N-m}}{2}\right).
\end{equation}

The strong-QD deficit plotted in Fig.~1 is
\begin{equation}
\Delta_{\rm SQD}(m)
=
H_\Pi-\chi_\Pi(m)+D_\Pi(m),
\end{equation}
that is,
\begin{equation}
\Delta_{\rm SQD}(m)
=
1
-
h_2\!\left(\frac{1+s^m}{2}\right)
+
h_2\!\left(\frac{1+s^N}{2}\right)
-
h_2\!\left(\frac{1+s^{N-m}}{2}\right).
\end{equation}
The horizontal axis in Fig.~1 is the fragment fraction
\begin{equation}
f=\frac{m}{N},\qquad m=0,1,\ldots,N .
\end{equation}
For the displayed figure we used
\begin{equation}
N=80,\qquad s=0.78 .
\end{equation}
The shaded observer window is the set of fragment sizes satisfying
\begin{equation}
\frac{\Delta_{\rm SQD}(m)}{H_\Pi}<0.05 .
\end{equation}
Since \(H_\Pi=1\) in this binary equal-prior example, this condition is simply
\begin{equation}
\Delta_{\rm SQD}(m)<0.05 .
\end{equation}

This finite-product construction is not an additional assumption in the
general theorems.  It is only a minimal exactly solvable instance used to draw
the schematic window in Fig.~1.  In this model the left edge of the window is
controlled by the decay of the record error, equivalently by the growth of
\(\chi_\Pi(m)\), while the right edge is controlled by the decay of residual
coherence in the complement, equivalently by the growth of \(D_\Pi(m)\) as
\(m\) approaches \(N\).

\section{Asymmetry bound and achievable redundancy capacity}
\label{app:asymmetry-capacity}

This appendix proves the three claims in Theorem~\ref{thm:asymmetry-capacity}:
the single-fragment asymmetry bound, the global redundancy converse, and the
i.i.d. achievability statement.

\subsection{Single-fragment asymmetry bound}

For the Abelian phase monitor of Theorem~\ref{thm:asymmetry-capacity},
\begin{equation}
U_{F,x}
=
\exp\!\left[
-\ii\sum_{k\in F}\sum_{\mu=1}^{r_G}
\theta_{k,x}^{\mu}G_k^\mu
\right],
\label{eq:app-controlled-phase-unitary}
\end{equation}
and
\begin{equation}
\sigma_{F|x}=U_{F,x}\sigma_FU_{F,x}^\dagger.
\label{eq:app-phase-orbit}
\end{equation}
All conditional states have the same entropy.  Therefore the pointer Holevo
information is
\begin{align}
\chi_\Pi(S:F)
&=
S\!\left(\sum_xp_x\sigma_{F|x}\right)
-
\sum_xp_xS(\sigma_{F|x})
\nonumber\\
&=
S(\bar\sigma_F)-S(\sigma_F),
\label{eq:app-chi-orbit}
\end{align}
where
\begin{equation}
\bar\sigma_F=
\sum_xp_xU_{F,x}\sigma_FU_{F,x}^\dagger .
\label{eq:app-average-orbit-state}
\end{equation}

Let
\(\boldsymbol G_F=\{G_k^\mu:k\in F,\ \mu=1,\ldots,r_G\}\), and let
\(\Delta_{\boldsymbol G_F}\) be the pinching onto the joint eigenspaces of this
commuting local-generator family.  If \(P_{\boldsymbol g_F}\) is one of these
joint eigenspace projectors, then
\begin{equation}
U_{F,x}P_{\boldsymbol g_F}
=
\exp[-\ii\Phi_x(\boldsymbol g_F)]P_{\boldsymbol g_F},
\qquad
\Phi_x(\boldsymbol g_F)=
\sum_{k\in F}\sum_{\mu=1}^{r_G}\theta_{k,x}^{\mu}g_k^\mu .
\end{equation}
Therefore the phase cancels inside each pinched block,
\(P_{\boldsymbol g_F}U_{F,x}\sigma_FU_{F,x}^\dagger P_{\boldsymbol g_F}
=P_{\boldsymbol g_F}\sigma_FP_{\boldsymbol g_F}\), and hence
\begin{align}
\Delta_{\boldsymbol G_F}(\bar\sigma_F)
&=
\sum_xp_x\Delta_{\boldsymbol G_F}
\!\left(U_{F,x}\sigma_FU_{F,x}^\dagger\right)
\nonumber\\
&=
\sum_xp_x\Delta_{\boldsymbol G_F}(\sigma_F)
=
\Delta_{\boldsymbol G_F}(\sigma_F).
\label{eq:app-dephase-average}
\end{align}
Pinching cannot decrease entropy, so
\begin{equation}
S(\bar\sigma_F)
\le
S(\Delta_{\boldsymbol G_F}\bar\sigma_F)
=
S(\Delta_{\boldsymbol G_F}\sigma_F).
\label{eq:app-dephasing-entropy}
\end{equation}
Substituting this into Eq.~\eqref{eq:app-chi-orbit} gives
\begin{align}
\chi_\Pi(S:F)
&\le
S(\Delta_{\boldsymbol G_F}\sigma_F)-S(\sigma_F)
\nonumber\\
&=
A_{\boldsymbol G_F}(\sigma_F),
\label{eq:app-asym-bound}
\end{align}
which is Eq.~\eqref{eq:asym-bound-main}.  The local-family pinching is essential:
coarse dephasing by the total charge \(G_F\) alone can leave coherences inside
degenerate total-charge sectors and therefore is not a stable resource account
for disjoint local records.

\subsection{Redundancy converse}

Let \(F_1,\ldots,F_R\) be disjoint fragments.  Applying
Eq.~\eqref{eq:app-asym-bound} to each fragment gives
\begin{equation}
\sum_{i=1}^{R}\chi_\Pi(S:F_i)
\le
\sum_{i=1}^{R}A_{\boldsymbol G_{F_i}}(\sigma_{F_i}).
\label{eq:app-sum-chi-local-asym}
\end{equation}
Let \(F_{\rm all}=\cup_iF_i\).  The joint local-generator family on
\(F_{\rm all}\) is \(\boldsymbol G_{F_{\rm all}}=\{G_k^\mu:k\in F_{\rm all},\ \mu=1,\ldots,r_G\}\), and
\(\Delta_{\boldsymbol G_{F_{\rm all}}}\) acts locally across the disjoint
fragments.  A direct subtraction gives
\begin{align}
&A_{\boldsymbol G_{F_{\rm all}}}(\sigma_{F_{\rm all}})
-
\sum_{i=1}^{R}A_{\boldsymbol G_{F_i}}(\sigma_{F_i})
\nonumber\\
&\quad=
\left[\sum_iS(\sigma_{F_i})-S(\sigma_{F_{\rm all}})\right]
-
\left[\sum_iS(\Delta_{\boldsymbol G_{F_i}}\sigma_{F_i})
-S(\Delta_{\boldsymbol G_{F_{\rm all}}}\sigma_{F_{\rm all}})\right].
\label{eq:app-asym-superadd-step}
\end{align}
The first bracket is the total mutual information among the fragments before
local-generator dephasing; the second bracket is the same quantity after the
local dephasing channel.  Local quantum channels cannot increase total mutual
information, so
\begin{equation}
A_{\boldsymbol G_{F_{\rm all}}}(\sigma_{F_{\rm all}})
\ge
\sum_{i=1}^{R}A_{\boldsymbol G_{F_i}}(\sigma_{F_i}).
\label{eq:app-asym-superadd}
\end{equation}
Partial trace is covariant with respect to the local phase symmetry and cannot
increase the relative entropy of asymmetry.  Therefore
\begin{equation}
A_{\boldsymbol G_{F_{\rm all}}}(\sigma_{F_{\rm all}})
\le
A_{\boldsymbol G_E}(\sigma_E),
\qquad
\boldsymbol G_E=\{G_k^\mu:k=1,\ldots,N,\ \mu=1,\ldots,r_G\}.
\label{eq:app-asym-partial-trace}
\end{equation}
Combining Eqs.~\eqref{eq:app-sum-chi-local-asym},
\eqref{eq:app-asym-superadd}, and \eqref{eq:app-asym-partial-trace},
\begin{equation}
\sum_{i=1}^{R}\chi_\Pi(S:F_i)
\le
A_{\boldsymbol G_E}(\sigma_E).
\label{eq:app-sum-chi-global-asym}
\end{equation}

If every counted fragment is a \(\delta\)-accurate strong-QD witness,
\begin{equation}
\sqdd(S:F_i)\le\delta,
\label{eq:app-delta-objective}
\end{equation}
then Eq.~\eqref{eq:sqddef-main} implies
\begin{equation}
\chi_\Pi(S:F_i)
\ge H_\Pi-\delta.
\label{eq:app-objective-record-lower}
\end{equation}
Consequently, for the maximum number \(R_\delta(N)\) of disjoint such fragments,
\begin{equation}
R_\delta(N)[H_\Pi-\delta]_+
\le
A_{\boldsymbol G_E}(\sigma_E),
\label{eq:app-red-converse}
\end{equation}
which is the global redundancy converse.  If the environmental input is a
product state, \(\sigma_E=\bigotimes_{k=1}^N\sigma_k\), then the local-family
asymmetry is additive:
\begin{equation}
A_{\boldsymbol G_E}(\sigma_E)
=
\sum_{k=1}^N A_{\boldsymbol G_k}(\sigma_k).
\label{eq:app-product-asym-additive}
\end{equation}
This gives Eq.~\eqref{eq:red-product-converse-main}.

\subsection{Achievability and capacity}

Assume a homogeneous i.i.d. exact monitoring stream,
\(\sigma_E=\sigma^{\otimes N}\), and a positive surplus-dephasing exponent for
the complement of every fixed-size block.  Equivalently, for each fixed block
size \(m\) there are constants \(b_m,c_m>0\) such that, for every block
\(|F|=m\),
\begin{equation}
R_{\rm dec}(E\setminus F)
\le b_m e^{-c_m(N-m)}.
\label{eq:app-positive-surplus-exponent}
\end{equation}
Let \(\chi_m\) be the pointer Holevo information of any block of \(m\) carriers,
and define
\begin{equation}
\mrec(\eta)=
\min\{m\ge1:\chi_m\ge H_\Pi-\eta\}.
\label{eq:app-mrec}
\end{equation}

The converse is a packing bound.  If \(F_i\) is counted by \(R_\delta(N)\), then
Eq.~\eqref{eq:app-objective-record-lower} gives
\(\chi_\Pi(S:F_i)\ge H_\Pi-\delta\).  By the i.i.d. definition of
\(\mrec(\delta)\), the fragment must contain at least \(\mrec(\delta)\) carriers.
Disjointness therefore implies
\begin{equation}
R_\delta(N)
\le
\left\lfloor\frac{N}{\mrec(\delta)}\right\rfloor,
\label{eq:app-capacity-converse-size}
\end{equation}
and hence
\begin{equation}
\limsup_{N\to\infty}\frac{R_\delta(N)}{N}
\le
\frac{1}{\mrec(\delta)}.
\label{eq:app-capacity-limsup}
\end{equation}

For achievability, fix \(0<\eta<\delta\) and partition the first
\(\lfloor N/\mrec(\eta)\rfloor\mrec(\eta)\) carriers into consecutive blocks of
size \(\mrec(\eta)\), ignoring the leftover carriers.  Each block satisfies
\begin{equation}
H_\Pi-\chi_\Pi(S:F_i)
\le\eta.
\label{eq:app-block-record-good}
\end{equation}
Its complement contains \(N-\mrec(\eta)\) carriers, so Eq.~\eqref{eq:app-positive-surplus-exponent}
and the continuity bound \eqref{eq:app-D-upper-R} give
\begin{equation}
D_\Pi(S:F_i)\to0
\qquad(N\to\infty)
\label{eq:app-block-discord-vanishes}
\end{equation}
uniformly over the blocks in the partition.  Therefore, for all sufficiently
large \(N\),
\begin{equation}
\sqdd(S:F_i)
\le\delta
\label{eq:app-block-objective}
\end{equation}
for every block in the partition, and hence
\begin{equation}
R_\delta(N)
\ge
\left\lfloor\frac{N}{\mrec(\eta)}\right\rfloor
\qquad(N\ \text{sufficiently large}).
\label{eq:app-capacity-achievability-size}
\end{equation}
Taking \(N\to\infty\) gives
\begin{equation}
\liminf_{N\to\infty}\frac{R_\delta(N)}{N}
\ge
\frac{1}{\mrec(\eta)}.
\label{eq:app-capacity-liminf}
\end{equation}
Together, Eqs.~\eqref{eq:app-capacity-limsup} and \eqref{eq:app-capacity-liminf}
give the sandwich bound for the capacity density.  If
\(\mrec\) is continuous at \(\delta\) as an integer-valued step function, then
letting \(\eta\uparrow\delta\) yields
\begin{equation}
\Cred(\delta)
=
\lim_{N\to\infty}\frac{R_\delta(N)}{N}
=
\frac{1}{\mrec(\delta)}.
\label{eq:app-capacity}
\end{equation}
At jump points one retains the one-sided liminf--limsup statement; the residual
mismatch is exactly the integer rounding of the minimal detector block size.

Finally, suppose the controlled-phase orbit converts local asymmetry into pointer
Holevo information without asymptotic loss,
\begin{equation}
\chi_m=
\min\{H_\Pi,mA_{\boldsymbol G}(\sigma)\}
\label{eq:app-asym-saturating}
\end{equation}
up to vanishing finite-size tolerances, where \(\boldsymbol G\) is the single-carrier generator family.  Then
\begin{equation}
\mrec(\delta)=
\left\lceil\frac{H_\Pi-\delta}{A_{\boldsymbol G}(\sigma)}\right\rceil,
\label{eq:app-mrec-saturating}
\end{equation}
and
\begin{equation}
\Cred(\delta)=
\left[
\left\lceil\frac{H_\Pi-\delta}{A_{\boldsymbol G}(\sigma)}\right\rceil
\right]^{-1}.
\label{eq:app-capacity-rounded}
\end{equation}
Thus
\begin{equation}
\Cred(\delta)
\le
\frac{A_{\boldsymbol G}(\sigma)}{H_\Pi-\delta},
\label{eq:app-density-asym-converse}
\end{equation}
and asymmetry-saturating monitors attain this converse up to the integer block-size
rounding in Eq.~\eqref{eq:app-capacity-rounded}.

\section{Tilted thermal spin-phase bath}
\label{app:spinphase}

This appendix derives Eqs.~\eqref{eq:spin-cdec}--\eqref{eq:spin-capacity}.
The example is a repeated-collision spin bath: the carrier state is prepared by
a tilted thermal Hamiltonian, and the monitoring collision itself is the Abelian
phase kick in Eq.~\eqref{eq:spin-phase-collision}.  This keeps the model inside
the controlled-product assumptions of Theorem~\ref{thm:sharp-window} and the
Abelian phase assumptions of Theorem~\ref{thm:asymmetry-capacity}.

Let
\begin{equation}
\boldsymbol n_0=(\sin\vartheta,0,\cos\vartheta),
\qquad
\sigma=\frac12(\one+r\boldsymbol n_0\cdot\boldsymbol\tau),
\qquad
r=\tanh\frac{\beta\Omega}{2}.
\label{eq:app-spinphase-sigma}
\end{equation}
When the temperature dependence is emphasized, we write this one-carrier state as
\(\sigma_T\).  The conditional phase rotations
\(U_\pm=\exp(\mp\ii\phi\tau_z/2)\) send the Bloch direction to
\begin{equation}
\boldsymbol n_\pm=
(\sin\vartheta\cos\phi,\ \pm\sin\vartheta\sin\phi,\ \cos\vartheta),
\qquad
\rho_\pm=\frac12(\one+r\boldsymbol n_\pm\cdot\boldsymbol\tau).
\label{eq:app-spinphase-conditional}
\end{equation}
The inhomogeneous case is obtained by adding a carrier index to
\(r,\vartheta,\phi\).

\subsection{Residual coherence}

For the binary pointer, the single-carrier contribution to the off-diagonal
system block is
\begin{align}
\gamma
&=\Tr(U_+\sigma U_-^\dagger)
 =\Tr(\sigma e^{-\ii\phi\tau_z})
\nonumber\\
&=
\Tr\!\left[
\frac12(\one+r\sin\vartheta\,\tau_x+r\cos\vartheta\,\tau_z)
(\cos\phi\,\one-\ii\sin\phi\,\tau_z)
\right]
\nonumber\\
&=
\cos\phi-\ii r\cos\vartheta\sin\phi .
\label{eq:app-spinphase-gamma}
\end{align}
Therefore
\begin{equation}
|\gamma|^2=
\cos^2\phi+r^2\cos^2\vartheta\sin^2\phi,
\qquad
c_{\rm dec}=-\log|\gamma|^2.
\label{eq:app-spinphase-cdec}
\end{equation}
For a complement \(\bar F\), the product form gives
\begin{equation}
|\Gamma_{\bar F}|^2=\prod_{k\in\bar F}|\gamma_k|^2,
\qquad
C_{\rm dec}(\bar F)=\sum_{k\in\bar F}c_{{\rm dec},k}.
\label{eq:app-spinphase-Cdec}
\end{equation}
This is the surplus-dephasing edge associated with Eq.~\eqref{eq:spin-cdec}.

\subsection{Record Chernoff exponent}

The two conditional states have the same eigenvalues
\(\lambda_\pm=(1\pm r)/2\).  For equal-spectrum qubit states related by a
reflection \(\phi\mapsto-\phi\), the Chernoff function is symmetric in
\(s\leftrightarrow1-s\) and is minimized at \(s=1/2\).  Write
\(\sqrt{\rho_\pm}=a\one+b\boldsymbol n_\pm\cdot\boldsymbol\tau\), where
\begin{equation}
a^2=\frac{1+\sqrt{1-r^2}}{4},
\qquad
b^2=\frac{1-\sqrt{1-r^2}}{4}.
\end{equation}
Then
\begin{align}
q
&=\inf_{0\le s\le1}\Tr(\rho_+^s\rho_-^{1-s})
=\Tr(\sqrt{\rho_+}\sqrt{\rho_-})
\nonumber\\
&=2(a^2+b^2\boldsymbol n_+\cdot\boldsymbol n_-)
=1-\left(1-\sqrt{1-r^2}\right)\sin^2\vartheta\sin^2\phi .
\label{eq:app-spinphase-q}
\end{align}
Thus \(c_{\rm rec}=-\log q\), and for a product fragment
\begin{equation}
C_{\rm rec}(F)=\sum_{k\in F}c_{{\rm rec},k}.
\label{eq:app-spinphase-Crec}
\end{equation}
This is the record edge associated with Eq.~\eqref{eq:spin-crec}.

\subsection{Local Abelian asymmetry and one-carrier Holevo information}

The local generator is \(G=\tau_z/2\).  Pinching in its eigenbasis removes the
transverse Bloch component:
\begin{equation}
\Delta_G(\sigma)=\frac12(\one+r\cos\vartheta\,\tau_z).
\end{equation}
Hence
\begin{equation}
A_G(\sigma)=S(\Delta_G\sigma)-S(\sigma)
=h_2\!\left(\frac{1+r\cos\vartheta}{2}\right)
-h_2\!\left(\frac{1+r}{2}\right).
\label{eq:app-spinphase-A}
\end{equation}
For equal pointer priors, the one-carrier average state is
\begin{equation}
\bar\rho_1=\frac{\rho_++\rho_-}{2}
=\frac12\left[
\one+r(\sin\vartheta\cos\phi\,\tau_x+\cos\vartheta\,\tau_z)
\right],
\end{equation}
whose Bloch radius is
\begin{equation}
\bar r_1
=r\sqrt{\cos^2\vartheta+\sin^2\vartheta\cos^2\phi}.
\end{equation}
Therefore
\begin{equation}
\chi_1
=h_2\!\left(\frac{1+\bar r_1}{2}\right)
-h_2\!\left(\frac{1+r}{2}\right)
\le A_G(\sigma),
\label{eq:app-spinphase-chi1}
\end{equation}
because \(\bar r_1\ge |r\cos\vartheta|\) and the binary entropy decreases as
the Bloch radius increases on \([0,1]\).  Equality holds for a relative
\(\pi\)-phase flip, \(\phi=\pi/2\), where the conditional ensemble performs the
local dephasing twirl exactly.  For arbitrary fragments, the general proof of
Theorem~\ref{thm:asymmetry-capacity} gives
\begin{equation}
\chi_\Pi(S:F)\le\sum_{k\in F}A_{G_k}(\sigma_k),
\label{eq:app-spinphase-asym-fragment}
\end{equation}
which is the fragment form of the main-text bound in Eq.~\eqref{eq:asym-bound-main}.

\subsection{Block Holevo information and redundancy capacity}

For a homogeneous stream, the Holevo information in a block of \(m\) carriers is
\begin{equation}
\chi_m
=
S\!\left(\frac{\rho_+^{\otimes m}+\rho_-^{\otimes m}}{2}\right)
-mh_2\!\left(\frac{1+r}{2}\right).
\label{eq:app-spinphase-chim}
\end{equation}
The numerical evaluation used for Fig.~\ref{fig:spin_bath}
is exact but low-dimensional.  Since \(\rho_\pm^{\otimes m}\) are permutation
invariant, Schur-Weyl decomposition gives
\begin{equation}
\rho_\pm^{\otimes m}
\cong
\bigoplus_j
\one_{\nu_j}\otimes
\frac{\exp(2\eta\boldsymbol n_\pm\cdot\boldsymbol J_j)}{(2\cosh\eta)^m},
\qquad
\eta=\operatorname{arctanh}r,
\label{eq:app-spinphase-schur}
\end{equation}
where \(\boldsymbol J_j\) are spin-\(j\) matrices and
\begin{equation}
\nu_j=
\binom{m}{m/2-j}-\binom{m}{m/2-j-1}
\label{eq:app-spinphase-multiplicity}
\end{equation}
with invalid binomial coefficients set to zero.  The largest block has dimension
\(m+1\), so evaluating \(\chi_m\) does not require diagonalizing a
\(2^m\)-dimensional matrix.

Given a target deficit \(\delta\), the operational record length is
\begin{equation}
m_{\rm rec}(\delta;T)=\min\{m:\chi_m(T)\ge1-\delta\}.
\label{eq:app-spinphase-mrec}
\end{equation}
The achievability part of Theorem~\ref{thm:asymmetry-capacity} partitions a long
stream into blocks of this size, while Theorem~\ref{thm:sharp-window} requires
the remaining complement to supply surplus dephasing.  Therefore, at continuity
points of \(m_{\rm rec}\),
\begin{equation}
\Cred(\delta;T)=\frac1{m_{\rm rec}(\delta;T)}.
\label{eq:app-spinphase-Cred}
\end{equation}
The asymmetry converse gives the density bound
\begin{equation}
\Cred(\delta;T)
\le
\frac{A_G(\sigma_T)}{1-\delta},
\label{eq:app-spinphase-Cred-bound}
\end{equation}
which is the upper curve in Fig.~\ref{fig:spin_bath}.

\subsection{High-temperature scaling and certified window}

For \(k_BT\gg\Omega\),
\begin{equation}
r=\tanh\frac{\beta\Omega}{2}\simeq\frac{\Omega}{2k_BT}.
\end{equation}
Expanding Eqs.~\eqref{eq:app-spinphase-cdec}, \eqref{eq:app-spinphase-q}, and
\eqref{eq:app-spinphase-A} gives
\begin{align}
c_{\rm rec}
&\simeq
\frac{r^2}{2}\sin^2\vartheta\sin^2\phi,
&
A_G(\sigma)
&\simeq
\frac{r^2\sin^2\vartheta}{2\ln2},
\nonumber\\
c_{\rm dec}
&=
-\log\left(\cos^2\phi+r^2\cos^2\vartheta\sin^2\phi\right)
\longrightarrow
-\log\cos^2\phi .
\label{eq:app-spinphase-highT}
\end{align}
Thus the record exponent and the witness capacity are suppressed as \(T^{-2}\),
while the dephasing exponent remains finite whenever \(|\cos\phi|<1\).  In a
homogeneous bath with \(|F|=m\) and \(N\) total carriers,
\begin{equation}
C_{\rm rec}(F)=mc_{\rm rec}(T),
\qquad
C_{\rm dec}(\bar F)=(N-m)c_{\rm dec}(T).
\end{equation}
Substitution into Theorem~\ref{thm:sharp-window} gives the finite-temperature
observer window
\begin{equation}
 m\ge\frac{\log(1/\eps)+\log\log(1/\eps)+\cO(1)}{c_{\rm rec}(T)},
\qquad
 N-m\ge\frac{\log(1/\eps)+\cO(1)}{c_{\rm dec}(T)},
\end{equation}
which is the finite-temperature specialization of Eq.~\eqref{eq:m_window}.

\end{document}